# Ohmic Contact Formation Between

# Metal and AlGaN/GaN Heterostructure via Graphene Insertion

Pil Sung Park[1], Kongara M. Reddy[2], Digbijoy N. Nath[1], Zhichao Yang[1],

Nitin P. Padture[3], and Siddharth Rajan[1,2,*]

[1]Department of Electrical Engineering, The Ohio State University, Columbus, OH 43210, USA

[2]Department of Materials Science & Engineering, The Ohio State University, Columbus, OH 43210, USA

[3]School of Engineering, Brown University, Providence, RI 02912, USA

**Abstract**

A simple method for the creation of Ohmic contact to 2-D electron gas (2DEG) in AlGaN/GaN high electron-mobility transistors (HEMTs) using Cr/Graphene layer is demonstrated. A weak temperature dependence of this Ohmic contact observed in the range 77 to 300 K precludes thermionic emission or trap-assisted hopping as possible carrier-transport mechanisms. It is suggested that the Cr/Graphene combination acts akin to a doped *n*-type semiconductor in contact with AlGaN/GaN heterostructure, and promotes carrier transport along percolating Al-lean paths through the AlGaN layer. This new use of graphene offers a simple and reliable method for making Ohmic contacts to AlGaN/GaN heterostructures, circumventing complex additional processing steps involving high temperatures. These results could have important implications for the fabrication and manufacturing of AlGaN/GaN-based microelectronic and optoelectronic devices/sensors of the future.

\* Electronic mail: rajan@ece.ohio-state.edu

The high sheet-carrier-density 2-dimensional electron gas (2DEG) at the AlGaN/GaN heterointerface combined with the wide band gap, high breakdown voltage, and high carrier mobility, make AlGaN/GaN-based high electron-mobility transistors (HEMTs) suitable for high-temperature[1], high-frequency[2-5], and high-power[6-8] applications. Also, AlGaN/GaN-based optoelectronic devices have applications as emitters in the ultraviolet[9-13], near-infrared[14,15], mid-infrared[16], and terahertz[17-19] ranges. However, Ohmic contacts to AlGaN/GaN heterostructures require highly tailored recipes for multi-layer metallization (typically Ti/Al/Ni/Au) and annealing at high temperatures (typically above 800 °C)[20,21] due to the large barrier height associated with the metal/AlGaN/GaN interface. Such Ohmic contacts formed by metal spiking into the GaN layer have uncontrolled, poorly understood morphologies[20,21], and they are known to reduce device reliability. This also limits the minimum feature sizes that can be achieved in devices, which in turn limits their high-frequency performance. In addition, the high-temperature annealing steps preclude the use of some advanced fabrication approaches, such as gate-first, self-aligned processing. While regrown contacts[22,23] and implantation[24] provide an alternative, these approaches also add to the complexity of the processing.

The isolation of single-layer graphene[25,26] has led to the demonstration of several exciting properties in this 2-D crystal of carbon, such as ballistic conduction by massless Dirac fermions[27,28] quantum Hall effect[27-29], and size-dependent band gap[30]. Developments in the synthesis of high-quality, large-scale graphene by epitaxial growth[31] and chemical vapor deposition (CVD)[32], has expanded the choice of substrates for the fabrication of graphene-based devices[26]. Moreover, CVD graphene is now being incorporated into other devices[33]. In this letter we show that insertion of single-layer CVD graphene between Cr and AlGaN/GaN layers offers a simple and reliable way to create Ohmic contacts to AlGaN/GaN heterostructures without the need for high-temperature annealing or other additional processing steps.

In this study, 31 nm $Al_{0.28}Ga_{0.78}N$/GaN heterostructures were grown epitaxially on Fe-doped, semi-insulating GaN buffer layer on sapphire substrates (Lumilog, Vallauris, France) at ~770 ºC using plasma-assisted molecular beam epitaxy (MBE; Veeco Gen 930, Plainview, NY) under Ga-rich conditions[34]. The as-grown material was characterized using x-ray diffraction (XRD; BEDE D1, Durham, UK) ω-2θ scans (Fig. 1a), and the thicknesses and composition of the layers were determined using routine dynamical XRD simulation. Atomic force microscopy (AFM; Veeco DI 3100, Plainview, NY) measurements reveal a smooth surface, with expected step-flow growth. The root mean square (RMS) roughness was estimated to be less than 0.5 nm over a surface area of $5\times5$ $\mu m^2$ (Fig. 1a).

Single-layer graphene was grown by CVD on high-purity polycrystalline Cu foils (Alfa Aesar, Ward Hill, MA) of 25 μm thickness using a process described elsewhere[35]. Cleaned Cu foil pieces ($1\times1$ $cm^2$) were placed inside the CVD chamber consisting of a controlled-atmosphere quartz-tube furnace (Lindberg/Blue M, Asheville, NC), $CH_4+H_2$ CVD was performed at 1000 ˚C. The polymethyl methacrylate (PMMA) method[32] was used to transfer the CVD graphene from the Cu foil onto the as-grown AlGaN/GaN substrate. The transferred graphene was characterized by Raman spectroscopy (InVia Raman Microscope, Renishaw, Gloucestershire, UK) using a 514 nm wavelength, 1 mW laser. Figure 1b shows representative Raman spectrum from the transferred graphene, with the signature D-band, G-band, and 2D-band peaks at 1350 $cm^{-1}$, 1580 $cm^{-1}$, and 2700 $cm^{-1}$, respectively. The approximate 2D:G peak height ratio of 2:1 is indicative of single-layer graphene. The lower intensity D-band peak in the spectrum indicates presence of a small amount of defects in the graphene.

Metal/Graphene/AlGaN/GaN diodes (Fig. 2), and reference metal/AlGaN/GaN Schottky diodes without the graphene, were fabricated. In both types of diodes, an electron-beam evaporated Cr/Au/Ni metal stack (referred to as Cr) was used for contacts. In the case of the Cr/AlGaN/GaN Schottky diodes, the graphene layer was removed by $O_2$ plasma reactive ion etching (RIE) before metal evaporation. In both types of diodes, Ohmic contact to the 2DEG at the AlGaN/GaN interface was formed at the edge of the samples using pressed indium metal.

Current density-voltage (*I-V*) characteristics of Cr/Graphene/AlGaN/GaN and reference Cr/AlGaN/GaN diodes were measured using a semiconductor parameter analyzer (Agilent B1500 A, Santa Clara, CA) (Fig. 3). The reference Cr/AlGaN/GaN diode shows Schottky behavior, with several orders of magnitude rectification, which is to be expected[34]. In contrast, linear Ohmic-like behavior is observed in the Cr/Graphene/AlGaN/GaN diode, with an extracted resistivity of ~2 m$\Omega$.cm$^{-2}$. To confirm that the measured *I-V* characteristics are not related to room-temperature thermionic emission or trap-assisted hopping transport, *I-V* characteristics of both Cr/AlGaN/GaN and Cr/Graphene/AlGaN/GaN diodes were measured at low temperatures (77 K to 300 K range) in vacuum (~2×10$^{-6}$ Torr) (Figs. 4a and 4b). The Cr/AlGaN/GaN diode is found to be rectifying at all temperatures (Fig. 4a), with several orders of magnitude change in current density, which is consistent with the thermionic nature of carrier transport. The Cr/Graphene/AlGaN/GaN diode *I-V* characteristics (Fig. 4b) are Ohmic at all temperatures, with higher current density at lower temperatures due to reduced series resistances and higher electron mobilities. In particular, no thermally-activated transport mechanisms are evident from these measurements, eliminating thermionic emission or trap-assisted hopping transport as possible reasons for the observed Ohmic behavior.

While the specific resistance of 2 m$\Omega$.cm$^{-2}$ is higher than the state-of-the-art, to the best our knowledge there are no other published reports showing a metal stack by itself making Ohmic contact to a 2DEG through a thick AlGaN barrier. Optimization of graphene quality, metal-stack parameters, and the epitaxial AlGaN/GaN structure could lead to further reduction in the specific resistances, with important implications for making contact in future group III nitride devices.

Chromium, with work function $\Phi_{Cr}$=4.28 eV was chosen as the metal in contact with graphene because it pins the work function of graphene to the value $\Phi_{Cr/Graphene}$=4.28 eV[37]. Cr in contact with bare AlGaN results in a Schottky barrier $\Phi_B$~0.8 eV, and based on the effective work function, would be expected to be the same in the case of Cr/Graphene/AlGaN. While *I-V* characteristics of Graphene/AlGaN have not been reported, they have been reported for graphene/GaN, and show Schottky behavior with a $\Phi_B$~0.74 eV[38]. Thus, it is likely that graphene alone would also provide Schottky contact at the AlGaN surface. Thus, while Cr, or graphene, alone result in Schottky contacts, Cr/Graphene combination provides Ohmic contact with AlGaN/GaN.

In order to analyze these results, first consider the Cr/AlGaN/GaN junction. In forward bias, the flow of electrons from the 2DEG to the metal takes place with a threshold of ~0.3 V (Fig. 4a). For metal/AlGaN/GaN junctions in general it is observed that that onset of current from the 2DEG to the metal takes place when there is no opposing field in the AlGaN. We attribute this to percolation transport of carriers from the GaN layer into Al-lean (GaN-like) regions in the AlGaN layer, which has been observed in GaN/AlGaN/GaN heterostructures previously[39]. In metal/AlGaN/GaN contacts in general, this results in conduction between the 2DEG and gate under flat band conditions for the AlGaN layer, when there is no electrostatic barrier to percolative transport for carriers in AlGaN. Thus, for Cr/AlGaN with $\Phi_B \sim 0.8$ eV, and a conduction band offset of 0.45 eV, the measured ~0.3 V turn-on voltage for the Cr/AlGaN/GaN Schottky junction is reasonable. The "effective" energy band diagram for this situation is shown in Fig. 5a, where the top and the bottom diagrams depict zero-bias and forward-bias conditions, respectively. The more widely studied Ni/AlGaN/GaN Schottky show the same behavior, as the gate turn-on is usually at a gate bias of +0.8 ~ 1 V[40] when the field in the AlGaN reaches flat band. In reverse bias, the metal/semiconductor barrier between metal and GaN prevents electron transport, although the relatively weak temperature dependence (~10-fold increase in current density over a temperature difference of 223 K; Fig. 4a) can be attributed to the low effective barrier height ($\Phi_B \sim 0.2$ eV) for transport from Cr to the GaN. In the case of Ni/AlGaN Schottky diodes[40], as expected the reverse bias leakage depends on the barrier height of the Ni/GaN junction (0.81 eV), rather than that of the Ni/AlGaN junction (1.5-1.6 eV). Thus, while the electrostatic barrier height for the heterostructure is set by the AlGaN alloy, there is a lower (percolation) barrier to transport in either direction.

In the case of Cr/Graphene/AlGaN/GaN, the *I-V* characteristics are symmetric, and there is no threshold for conduction (Fig. 4b). The absence of a threshold voltage for current in forward- and reverse-bias conditions indicates zero-field in the AlGaN, in contrast to the Cr/AlGaN/GaN case. The energy band diagram is, therefore, expected to be symmetric (Fig. 5b), with the Cr/graphene providing the same energy line-up relative to the AlGaN as the underlying GaN layer. The effective conduction band for percolation (Fig. 5b), therefore, presents no barrier to transport in forward or reverse bias. This model is consistent with previous work, where Cr is shown to interact with graphene strongly, relative to other metals such as Au and Fe[41], and open a band gap in Cr/Graphene making it behave like a semiconductor[42]. It is suggested that the Cr/Graphene combination here behaves akin to doped n-GaN. While further work is needed to validate this hypothesis, an important conclusion of this model is that thinner AlGaN layers are likely to lead to significantly lower contact resistance because the probability of percolation through the AlGaN is likely to increase.

In closing, we have shown that the insertion of single-layer graphene between Cr metal layer and AlGaN/GaN semiconductor heterostructure provides an Ohmic contact with a specific resistance of ~ 2 m$\Omega$.cm$^{-2}$. It is proposed that the Cr/Graphene combination behaves akin to a doped n-type semiconductor leading to a symmetric energy line-up, and that percolation of Al-lean (GaN-like) regions promotes carrier transport through the AlGaN layer, leading to the Ohmic behavior. This new method of making Ohmic contacts in GaN-based HEMTs could provide a superior alternative to high-temperature annealed or regrown contacts. The results reported here could have important implications for the fabrication and manufacturing of AlGaN/GaN-based microelectronic and optoelectronic devices/sensors of the future.

**Acknowledgements**

The authors thank Prof. W. Windl for fruitful discussions. Funding for this work was provided by the National Science Foundation (grant no. NSF ECS-0925529) and the Office of Naval Research DATE MURI (grant no. ONR N00014-11-1-0721, Program Manager: Dr. Paul Maki).

**Figure Captions**

Figure 1. (a) Measured XRD ω-2θ scan and corresponding fit from as-grown AlGaN/GaN structure. Inset: schematic diagram of the heterostructure stack (UID refers to unintentionally doped; not to scale) and AFM image of surface morphology. (b) Raman spectrum from transferred CVD graphene on AlGaN/GaN substrate (inset), with D, 2D, and G bands indicated.

Figure 2. Schematic diagram showing the structure of the Cr/Graphene/AlGaN/GaN structure. In the case of the reference device, the graphene layer is removed. Not to scale.

Figure 3. Current-voltage characteristics of Cr/AlGaN/GaN and Cr/Graphene/AlGaN/GaN.

Figure 4. Current-voltage characteristics as a function of temperature (77 to 300 K, $2\times10^{-6}$ Torr) of: (a) Cr/AlGaN/GaN and (b) Cr/Graphene/AlGaN/GaN.

Figure 5. Schematic energy band diagrams of: (a) Cr/AlGaN/GaN Schottky diode under zero-bias (top) and forward-bias (bottom) conditions, and (b) Cr/Graphene/AlGaN/GaN junction (proposed). $E_C$ and $E_F$ indicate the bottom of the conduction band and the Fermi level, respectively. Dashed lines indicate effective $E_C$ for percolative transport.

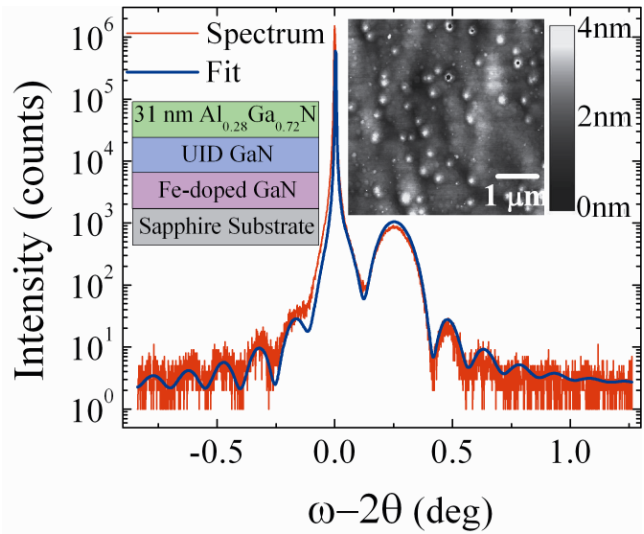

Figure 1 (a)

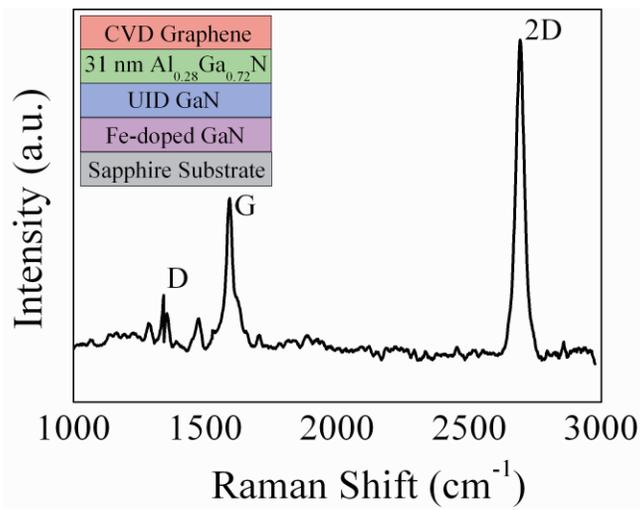

Figure 1 (b)

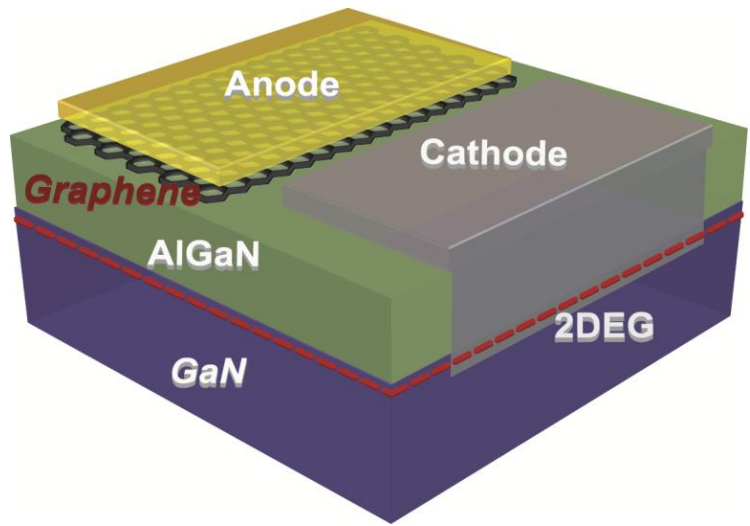

Figure 2

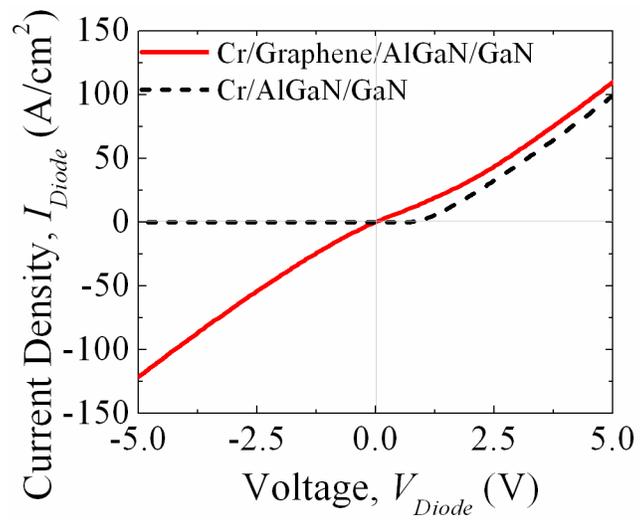

Figure 3

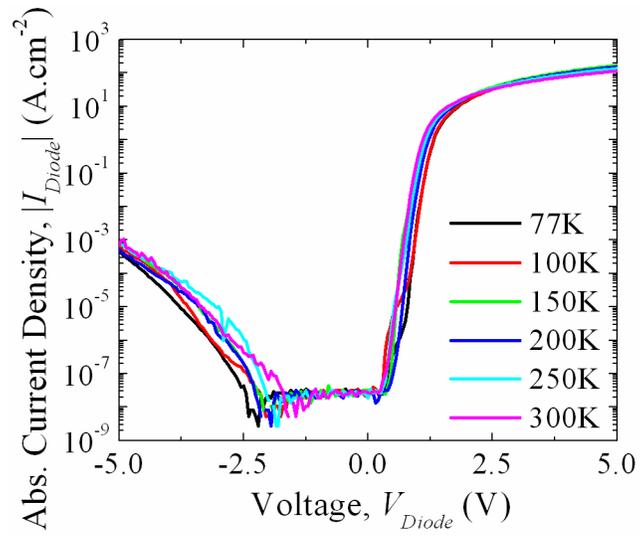

Figure 4 (a)

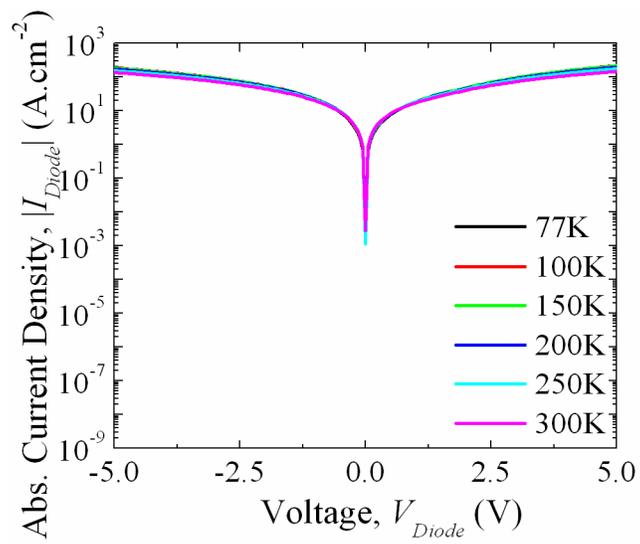

Figure 4 (b)

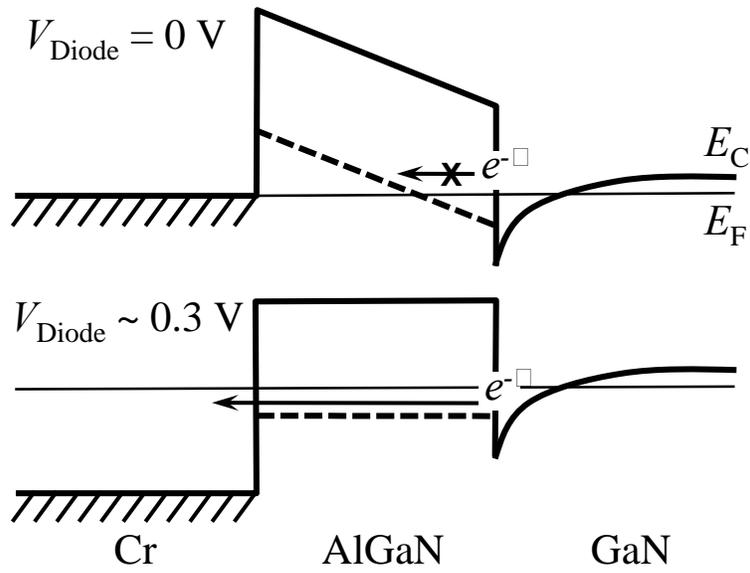

Figure 5 (a)

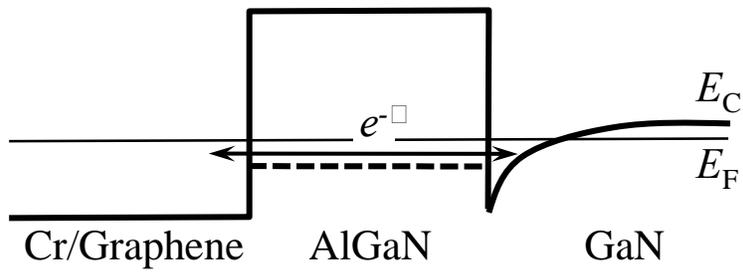

Figure 5 (b)